\documentclass[12pt,preprint]{aastex}
\usepackage{emulateapj5}
\usepackage{natbib,apjfonts,graphicx}

\shorttitle{High Resolution X-ray Spectroscopic Constraints on Cooling-Flow Models}
\shortauthors{Peterson et al.}

\begin{document}

\journalinfo{}

\title{High Resolution X-ray Spectroscopic Constraints on Cooling-Flow Models for Clusters of Galaxies}

\author{J. R. Peterson, S. M. Kahn, F. B. S. Paerels}
\affil{Columbia University, Columbia Astrophysics Laboratory, 538 W. 120th St., New York, NY 10027, USA}
\email{jrpeters@astro.columbia.edu, skahn@astro.columbia.edu, frits@astro.columbia.edu}
\author{J. S. Kaastra, T. Tamura, J.~A.~M.~Bleeker, C. Ferrigno}
\affil{SRON National Institute for Space Research, Sorbonnelaan 2, 3584 CA Utrecht, The Netherlands}
\email{J.S.Kaastra@sron.nl, T.Tamura@sron.nl, J.A.M.Bleeker@sron.nl, ferrigno@pa.iasf.cnr.it}
\author{J.~G.~Jernigan}
\affil{Space Sciences Laboratory, University of California, Berkeley, CA 94720, USA}
\email{jgj@ssl.berkeley.edu}

\slugcomment{Submitted to ApJ}

\begin{abstract}
We present high resolution X-ray spectra of 14 putative cooling-flow clusters of galaxies obtained with the Reflection Grating Spectrometer on XMM-Newton.  The clusters in the sample span a large range of temperatures and mass deposition rates.  Various of these spectra exhibit line emission from O VIII, Ne X, Mg XII \& XI, Al XIII \& XII, Si XIV \& XIII, N VII, and C VI as well as all Fe L ions.  The spectra exhibit strong emission from cool plasma at just below the ambient temperature, $T_0$, down to $T_0/2$, but also exhibit a severe deficit of emission, relative to the predictions of the isobaric cooling-flow model at lower temperatures ($<$ $T_0/3$).  In addition, the best-resolved spectra show emission throughout the entire X-ray temperature range, but increasingly less emission at lower temperatures than the cooling-flow model would predict.

 These results are difficult to reconcile with simple prescriptions for distorting the emission measure distribution, e.g. by including additional heating or rapid cooling terms.  We enumerate some theoretical difficulties in understanding the soft X-ray spectra of cooling-flows independent of the classic problem of the failure to detect the cooling-flow sink.  Empirically, the differential luminosity distribution is consistent with being proportional to the temperature to the power of $\approx$ 1 to 2, instead of being independent of the temperature, as expected in the standard multi-phase model.  The primary differences in the observed low temperature spectra are ascribed to differences in the ambient temperature.
\end{abstract}

\section{Introduction}

It was recognized many years ago that the cores of clusters of galaxies have 
sufficient X-ray luminosity to cool 10 to 1000 solar masses of X-ray emitting
plasma every year  (e.g. \citealt{fabian2}, \citealt{cowie}, \citealt{fabian1}).  The
details of the cooling process are still debated, however, in most 
models, parcels of cooling plasma collect at the center
of the cluster (\citealt{nulsen1}), forming what is referred to as a cooling-flow.  

This basic picture has remained controversial, however, and a number of observational discrepancies and alternate theoretical interpretations have been raised in the literature.  While it is well-established that cores of clusters have short ($<$ $10^9$ yr) cooling times (e.g. \citealt{white2}, \citealt{peres}, \citealt{allen2}) and cluster cores have been shown conclusively to contain much lower X-ray temperature plasma than the ambient hot outer regions (\citealt{canizares1}, \citealt{canizares2}, \citealt{mushotzky}), there is no consensus about how much mass has cooled from X-ray temperatures.  Some evidence exists for copious amounts of cooler gas emitting in the UV (\citealt{oegerle}), and in H$\alpha$ emission (\citealt{heckman}, \citealt{crawford}).  In addition, molecular hydrogen has been detected (\citealt{jaffe}, \citealt{donahue}), and the existence of significant quantities of dust has been inferred from infrared emission (\citealt{edge2}, \citealt{irwin}, \citealt{allen}).  However, HI absorption measurements have found no evidence for cold condensed clouds (e.g. \citealt{odea}), and while CO emission has been detected (\citealt{edge}) the amount of cold molecular material is still a factor of 10 below what is predicted by X-ray cooling estimates.  Therefore, the connection between these observations and the X-ray data is still unclear, and the exact quantity and location of cooling-flow byproducts is unresolved.  We refer to this as the {\it classic cooling-flow problem}.

The bulk of the thermal energy of the cooling intracluster medium is thought to radiate at X-ray wavelengths, and thus studying the X-ray spectrum is critical to understanding cooling-flows and testing cooling-flow models.
If the plasma cools homogeneously, the density profile of the 
cores of clusters should be much steeper than observed (\citealt{johnstone}), which has led to the conclusion that the cooling plasma must condense locally into smaller clouds distributed over a large volume (tens of kpc), i.e. in a 
multi-phase medium.  However, even ignoring the details of the
resulting spatial distribution, simple thermodynamic arguments show that 
the integrated X-ray spectrum of such a cooling flow can be
robustly predicted.  If the blobs of plasma cool in thermal isolation at 
constant pressure, and the dominant energy loss mechanism is via X-radiation, then the differential luminosity distribution, i.e. the luminosity radiated per unit temperature interval, must be proportional to the mass deposition rate, $\dot{M}$:

\begin{equation}
\frac{dL}{dT} = \frac{5}{2} \frac{\dot{M} k}{\mu m_p}
\end{equation}

\noindent
where $k$ is Boltzmann's constant, and $\mu m_p$ is the mean mass per particle.  The only free parameter in this expression is $\dot{M}$, which 
can be estimated from the density distribution inferred from the X-ray image of the cluster core.  The resulting 
spectrum can be calculated using a collisional equilibrium spectral 
synthesis model with an assumed set of elemental abundances, and 
normalizing the contribution in each temperature interval as given in 
Equation (1).  To a good approximation, a cluster spectrum should consist of two components: 1) the cooling-flow spectrum, as described above, and 2) an isothermal spectrum evaluated at the temperature of the background cluster gas.

Data acquired by the Reflection Grating Spectrometer (RGS) on XMM-Newton 
enable this robust spectral prediction to be quantitatively tested for 
the first time.  Surprisingly, the observed spectra reveal a remarkable 
systematic deficit of emission at the lowest temperatures, as compared to the 
multi-phase model (\citealt{peterson1}, \citealt{tamura}, \citealt{kaastra}, \citealt{tamura2}, \citealt{xu}, \citealt{sakelliou}).  This result has been confirmed through medium resolution spatially-resolved 
spectroscopy using the XMM-Newton European Photon Imaging Cameras (EPIC) and Chandra ACIS observations, where spectral fits have yielded significantly smaller $\dot{M}$'s than expected
(\citealt{david}, \citealt{boehringer}, 
\citealt{molendi}, \citealt{schmidt}, 
\citealt{ettori}, \citealt{johnstone2}).   We refer to the observed deficit of the predicted soft X-ray emission as the {\it soft X-ray cooling-flow problem}.  As we discuss in this paper, the soft X-ray cooling-flow problem may or may not be related to the classic cooling-flow problem.

Here we present a systematic study of 14 clusters with the RGS to quantify in detail the soft X-ray cooling-flow problem.  We present four basic results:  1) We find no evidence for X-ray absorption by cold material, which had been inferred from lower spectral and angular resolution data \citep{white}.  2) We find a severe lack of emission from lower temperature ions expected in the standard cooling-flow model.  3) We demonstrate the ubiquity of significant plasma just below the ambient gas temperature, $T_0$, down to $T_0/2$ in roughly the predicted amount.  This last result is the most perplexing and is difficult to reconcile with proposed explanations for the cooling-flow problems.  4) We offer an empirical parameterization of the temperature distribution which is consistent with the entire sample.

The paper is organized as follows.  In \S2 we discuss the expected spectrum of the isobaric radiative multi-phase model and various associated temperature diagnostics.  In \S3 we discuss the spectral capabilities of the RGS for sources of moderate extent.  In \S4 we describe our analysis methods and present the high resolution spectra.  In \S5 we discuss several qualitative aspects of the spectra.  In \S6 we describe the model used to set cooling luminosity limits.  In \S7 we present the results.  In \S8 we discuss the results with respect to potential modifications of the cooling-flow scenario.


\section{Diagnostics of the Isobaric Radiative Multi-phase Model}

It has been argued that the intracluster medium (ICM) should remain in collisional ionization equilibrium, even as it cools.  Both the recombination time scale and electron equilibriation time scale through Coulomb collisions, are much less than the cooling time scale, evaluated at characteristic X-ray temperatures and densities in the ICM (\citealt{edgar}, \citealt{hicks}).  At high temperatures, the X-ray spectrum is produced through bremsstrahlung and line emission from abundant elements. The spectra are dominated by lines from hydrogen-like and helium-like ions, and iron L shell emission.   In principle, the temperature distribution can be determined by three different methods:  1) through the shape of the exponential cut-off of the bremsstrahlung spectrum, 2) the ratios of hydrogen-like to helium-like lines, and 3) through the distribution of line emission from iron L-shell ions.  However, the bremsstrahlung spectrum constrains only the highest temperatures present in the spectrum, and the ratio of hydrogen-like to helium-like lines is relatively unconstraining since both charge states are present over a very broad range of temperatures.  Therefore, Fe L emission is the most useful diagnostic, and provides the primary method for determining the temperature distribution presented in this paper.

The isobaric multi-phase model yields a unique spectral signature given an isothermal outer cluster temperature and a set of assumed abundances.  This is shown in Figure 1 for a maximum temperature of 8 keV and for 1/3 solar abundances.  In Figure 2, we divide the temperature distribution below 6 keV into four bands.  The curves are the spectra produced by emission from the multi-phase model in bins of 3 to 6, 1.5 to 3, 0.75 to 1.5, and 0.375 to 0.75 keV.  Important line blends are compiled in Table 1.  Several important diagnostics and aspects of the model can be noted.  The first is that line emission from ions between Fe XXIV and Fe XVII provide tight constraints on the temperature distribution.  There is roughly comparable emission in the emission line blends from each Fe L ion.  Also note that Fe XVII line emission is predicted to be three times as strong as Fe XVIII.  O VIII and other hydrogen-like ions are produced at a range of temperatures, and thus do not provide strong constraints on the temperature distribution.  The overall normalization is proportional to the mass deposition rate, and the spectrum should be a strict superposition of all the individual temperatures up to the ambient temperature.


\section{The Use of the Reflection Grating Spectrometers for Observations of Extended Sources}

In order to test aspects of this model in detail, the Fe L spectrum needs to be resolved for a spatially-extended object.  
This could not be accomplished with previous instrumentation; however, the launch of the XMM-Newton RGS experiment now makes it possible.  For a full description of the RGS experment see \cite{denherder}.
  The two RGS spectrometers have 160 $\mbox{cm}^2$ of combined collecting area and with a spectral resolution given roughly by: 

\begin{equation}
\Delta \lambda
\approx 0.12 {\mbox{\AA}} \times {\mbox{size~of~source~in~arcminutes}} / {\mbox{spectral~order}}
\end{equation}

\noindent
 for an extended source larger than 10 arcseconds.  The wavelength band extends from 5 to 38 ${\mbox{\AA}}$, which samples Si K shell transitions to C K transitions.  The field of view is effectively 5 arcminutes by 1 degree, which is large enough to capture the entire cooling-flow region. The former is set by the width of the CCD array in the cross-dispersion direction, and the latter is set by self-vignetting of the telescope shells.  For the cooling-flow clusters, data selection cuts are tailored to include mostly emission from the cores of the clusters.  The background is produced primarily by a variable flux of soft protons spread nearly uniformly across the focal plane.  There is no source-free region for extended sources, so it has to be modeled.  The cross-dispersion direction provides a one-dimensional image (11 arcseconds FWHM) of the source, which we use to construct a model of the surface brightness of the source.


\section{Sample Selection and Fluxed Spectra}

Our sample of clusters (listed in Table 2) is a biased selection of 14 compact clusters and groups of galaxies chosen to exploit the spectral sensitivity of the RGS.  The sample includes clusters and groups at a range of temperature from 1 to 10 keV.    All of these clusters and groups were expected to host cooling-flows ranging from 1 to 1000 solar masses per year (see Table 3).  

To produce effective-area and exposure-corrected spectra of these objects, we adopt the following procedure:  We process the events by a development version of the Science Analysis System (SAS) software (version 5.3), which accomplishes event reconstruction, aspect correction, CCD-pulseheight corrections, and focal plane reconstruction.  We only use events in time intervals where the background was less than 6 counts per second in the 0.35 to 1.9 keV energy band.
We then select photon events satisfying a 2 arcminute wide cross-dispersion cut.  Events must also satisfy a first or second order joint dispersion-coordinate/CCD-pulseheight cut based on the characteristic resolution of the CCD, and additional broadening due to the 2 arcminute source extent.  In this way, we select photons roughly produced in a 2 arcminute square centered on each cluster.  There is also a significant contribution of photons from outside this square, which is self-consistently in our modeling.  Finally, a wavelength is assigned to each photon based on the nominal center of the cluster and the instrument bore-sight.

In order to account for the instrument response and efficiency of these selection cuts, we use a Monte Carlo method, as discussed in detail by \cite{peterson3}.  The Monte Carlo method is used to account for off-axis behavior of the response and the effect of arbitrary selection cuts and transformations made on the data.  Additionally, it is required in the astrophysical modeling as we discuss in \S6.
  The Monte Carlo calculation contains all known effects of the mirror shells, grating arrays, and CCD response.  It has calibration limitations similar to the response in SAS version 5.3.  The Monte Carlo method generates focal plane coordinates and CCD-pulseheights based on a model for the X-ray spatial and spectral distribution.

 We generate photons having a flat wavelength spectrum and surface brightness profile given by a modified $\beta$ model (described in \S6).  The parameters are listed in Table 3.  The Monte Carlo method uses focal plane maps of the exposure time produced by the SAS to select detected events.  The same selection cuts that are applied to the data, are applied to the simulated photons as well to account properly for the efficiency of the selection regions.   Extracting the simulated photons as we did for the source photons produces a wavelength histogram in units of ${\rm cm}^2 {\rm s}~{\mbox{\AA}}$.  We then take the histogram of the source photons and subtract a set of background simulated events.  The background model is described in detail in \S6.  The final histogram is divided by the exposure-area histogram and produces the fluxed spectra shown in Figures 3a, 3b,  and 3c.  The spectra are not corrected for absorption in the interstellar medium.

\section{Qualitative Results on the Sample}

 The spectra shown in the three panels of Figure 3 are sorted inversely by total luminosity, and therefore also roughly inversely by temperature and $\dot{M}$.  Below, we discuss the spectra grouped by their outer temperature.

{\bf 7-10 keV Clusters}:  The RGS spectrum of Abell 1835 has already been presented in \cite{peterson1}.  The spectrum exhibits intermediate temperature ($k T_e$ $\approx$ 3 keV) plasma in addition to the 8 keV background plasma as shown by the detection of Fe XXIV and XXIII.   O VIII Ly$\alpha$ is clearly detected.  \cite{peterson1} identified the gross inconsistencies between the measured spectrum of this source and the spectrum predicted by the standard cooling-flow model.  These constraints are derived from the lack of Fe XVII-XXII.  Abell 665 has a similar spectrum, but the observation of that source was plagued by a very high background rate.

{\bf 4-7 keV Clusters}:  Abell 1795, Hydra A, Abell 496, and Abell 4059 are 4 to 7 keV clusters.  Fe XXIV-XXII, O VIII, Mg XII, Ne X, and Si XIV emission lines are detected in all cases.  The strong Fe XXIV-XXII complex indicates significant cooling down to $k{T_e}$ $\approx$ 2 to 3 keV.  There are no detections of Fe XVII-XXI, however, in clear contradiction to the predictions of the cooling-flow model for these systems.

{\bf 2-4 keV Clusters}:  The lowest temperature clusters, 2A0335+096, S\'{e}rsic 159-03, Abell 262, Abell 1837, Mkw3s, Abell 2052, and M87 have the bulk of their emission from the Fe L temperature range.  These spectra show clear detections Fe XXIV-XIX, O VIII, Mg XII, Ne X, Si XIV.  Various different ions produce stronger emission lines depending on the cluster temperature.  Weak emission from the helium-like charge states are observed for silicon and magnesium, and in the best resolved spectra hydrogen-like C and N are detected.  These spectra also generally have weak emission from Fe XVII and XVIII indicating some plasma below 1 keV, but less than would be predicted by the standard cooling-flow model.

{\bf 1 keV Groups}:  NGC 533 is the lowest temperature system and has emission from the same ions as in the 2-4 keV clusters.  In this case, Fe XVII and XVIII are strong due to the lower temperature of the system.  However, this spectrum is also inconsistent with the standard cooling-flow predictions, since Fe XVII is not much stronger than Fe XVIII and OVII is not detected.

{\bf Composite Spectrum}:  As further confirmation of the reality of some of the detections discussed above, we generate a composite spectrum by combining all of the individual clusters.  This composite is shown at the bottom of Figure 3.  It was constructed by first shifting the wavelengths of photons by the cosmological redshift and then co-adding the counts-weighted spectra.  The individual area-exposures were also added using the same technique.  The counts-weighted spectrum was then divided by the sum of the area-exposures.  We excluded M87 from this analysis since it would otherwise dominate the result.

In the composite spectrum, lines from Si XIV \& XIII, Al XIII \& XII, Mg XII \& XI, all Fe L ions, Ne X, O VIII, C VI, and N VII are all clearly visible.  No neutral O K edge is observed at the redshift of the clusters.  This would be the most prominent absorption feature if a large amount of cold material with a large covering fraction was absorbing the flow.  A feature which appears to be a blue-shifted O K absorption edge is due to galactic absorption, and its depth is consistent with the mean interstellar column density.

Several qualitative conclusions can be drawn from the observations:  First, in all cases, there is a deficit of emission expected from lower temperatures.  Second, in most cases, a continuous distribution of temperatures is nevertheless required.  Finally, absorption by intervening cool gas at the redshift of the cluster cannot explain the dearth of soft X-ray emission.  In \S 6 below, we perform quantitive spectral fits to further explore the departures from the cooling-flow model.

\section{Fitting Methods and Model}

In this section, we set quantitative limits on emission from various parts of the temperature distribution.  To derive these limits, we use the Monte Carlo methods presented by \cite{peterson3}, and implicitly used by \cite{peterson1} and \cite{xu}.  The basic procedure is based on an astrophysical model for the spatial and spectral dependence of the emission.  The Monte Carlo approach is required because a different spectrum at each projected spatial position is required to test the cooling-flow model in detail.  We randomly generate photons with an associated dispersion angle, cross-dispersion angle, and CCD pulseheight value.  The resulting count distributions are then compared with the raw data, after the various data selection cuts and transformations.  In fitting for global parameters, such as the background normalization, we find the best fitting solution by using multivariate methods, as described in \cite{peterson3}.  For the abundances and temperatures, we use a $\chi^2$ statistic of the combined extracted first and second order spectra for both instruments and we iteratively adjust the surface brightness distribution to match the cross-dispersion profile.

For the cluster emission we adopt a relatively simple model, so as to reduce sensitivity to fitting biases.  For the surface brightness, we use a spherical $\beta$ profile where the core radius is left free and the $\beta$ parameter is fixed to the value determined from EPIC spectral fits (\citealt{kaastra2}).  Additionally, we allow the normalization of the emission inside of a three-dimensional radius, $r_{cool}$, to be larger than the normalization outside of the radius.  In this way, the spatial profile can be much more peaked than the standard $\beta$ profile.  This three dimensional distribution is then projected on the sky.  The precise shape of the spatial distribution is not critically important for determining the cooling luminosity limits as long as it reproduces the rough behavior of the emission profile. 

  Outside of $r_{cool}$, the emission is set to an isothermal temperature.  Inside that radius, we fit for the normalization of the differential emission measure distribution below the upper temperature, $T_0$.  The emission measure is divided into several temperature bins between $\case{1}{2} T_0$ and $T_0$, $\case{1}{4} T_0$ and $\case{1}{2} T_0$, $\case{1}{8} T_0$ and $\case{1}{4} T_0$, and $\case{1}{16} T_0$ and $\case{1}{8} T_0$.  This choice is arbitrary, but it provides robust fitting solutions since the fractional ionization curves are roughly equally spaced in the logarithm of the temperature.  Preliminary fits indicate that only negligible hot ambient plasma coexists inside the cooling radius, so additional isothermal emission inside this radius is ignored.  Within each temperature bin, we use the isobaric radiative cooling-flow model to predict the shape of the emission measure distribution, but this is not critically important since we do not have the spectral sensitivity to sample the emission measure distribution in very fine intervals.  This approach clearly forces the coolest emission to conform to a specified spatial distribution.  Further detailed analyses are required to pinpoint its exact spatial location.  Our spatial model, however, seems to be compatible qualitatively with the relatively small differences in the observed emission lines spatial profiles of M87 (\citealt{sakelliou}).

We assume uniform spatial abundances throughout the cluster emission.  This considerably simplifies the fitting procedure, and makes the limits on the coolest emission conservative.  It is known that abundance gradients exist in clusters, but they are generally flat within the cooling-flow region and weaken whenever more freedom is given to the temperature distribution \citep{molendi}. The iron, neon, oxygen, magnesium, and silicon abundances are left as free parameters.  All other elements are tied to iron since they contribute few counts.   In Abell 1835 and Abell 665, the magnesium, neon, and silicon abundances are tied to iron since the emission is from very high temperatures.  The absorption column density is left as a free parameter to account for variations along the line of sight to the cluster.  In NGC 533 the absorption is set to the galactic value since it has little low energy continuum emission.  We also ignore the effects of resonant scattering (cf. \citealt{xu}), which redistribes emission line photons within our aperture, but otherwise results in the same detected emission line flux.

For the background, we use a semi-empirical model calibrated on blank sky Lockman Hole observations (XMM Revolutions 0070/0073).  The model includes a spatial model for soft protons, low energy detector readout noise, and characterizations of the in-flight Al K and F K calibration sources.  All parameters are frozen in this model except the relative normalization of the particle component, and the overall normalization which can vary by factors of 10 from observation to observation.  The background is relatively flat in wavelength.

The model has the following free parameters:  local column density, normalization of each part of the cooling flow emission measure distribution, abundances of magnesium, neon, silicon, oxygen, and iron, background temperature ($T_0$), particle background normalization, position of the source in the cross-dispersion direction, core radius, cooling radius, and overall normalization.  For each spectrum, we apply selection cuts discussed in \S4.  However, the background model parameters and position of the source are determined before applying any data selection cuts.

In addition to the model described above, we also set limits on a model wherein an intrinsic absorber embedded in the cooling-flow volume is invoked to suppress the expected soft emission.  If the absorber is evenly embedded, the transmission function is given by, $\left( 1 - e^{-\tau(E)} \right) /\tau(E)$, where $\tau(E)$ is the photoelectric optical depth as a function of energy, as opposed to the usual exponential form.  For this case, we otherwise take the isobaric cooling-flow emission measure distribution, without allowing any variation from one temperature bin to the next.

The errors are quoted at the 90$\%$ statistical confidence level for setting limits on the parameters.
The uncertainty in the wavelength scale of 8 ${\mbox{m\AA}}$ and the characteristic line spread function uncertainty of 5 ${\mbox{m\AA}}$ make negligible contributions to these errors for extended sources like clusters.  The effective area uncertainty is of order 10$\%$.  Additionally, there is known to be non-statistical noise at the 5$\%$ level of the source flux in the RGS spectra, caused by systematic dark current variations which can sometimes produce false features.  Either of these effects can produce correlated errors in adjacent wavelength channels.
  The global temperature, absorption column density, and overall normalization depend on many wavelength channels and are thus dominated by systematic uncertainties of this kind.  These parameters are therefore never quoted with uncertainties below 200 eV, $5\times 10^{19} \mbox{cm}^{-2}$, or 10$\%$ in the normalization, respectively, which would all modify the spectrum characteristically by $\sim$ 10$\%$.

We adopt the MEKAL plasma model \citep{mewe} as implemented in XSPEC v11 (\citealt{arnaud}) as the collisional ionization equilibrium model used to predict the spectrum.  The absorption cross-sections are taken from \cite{morrison}.  Abundances are quoted relative to \cite{anders}.  Galactic absorption column densities are compared to \cite{dickey}.  We assume ${\rm H_0}=70$ km/s/Mpc, $\Lambda=0.7$, $\Omega_m=0.3$ throughout our analysis.

Finally, all measured spectral mass deposition rates are compared to those derived in the corresponding EPIC data set.  That analysis takes into account temperature and density gradients by solving the hydrodynamic equations, assuming only radiative cooling and mass drop out according to standard techniques (e.g. \citealt{white2}).  The mass deposition rates are interpolated at the radius for our $r_{cool}$ parameter and are within uncertainties (typically 50$\%$) of other published values.  A more extensive discussion of the EPIC analysis will be presented elsewhere (\citealt{kaastra2}).

\section{Cooling Luminosity Limits}

Our derived spectral fits are shown in Figure 4a, 4b, and 4c and the cross-dispersion profiles are shown in Figure 5.  The best fit parameters are given in Tables 3, 4 and 5.  In each plot the red line shows the fitted empirical model.  The green line shows the standard isobaric cooling-flow model without altering the temperature distribution and maintaining the same normalization.  In almost all cases, the spectral features and the intrinsic line profiles are well-fit for the empirical model.  Due to the low signal to noise of the observation of Abell 1837 and Abell 665, we could not place strong constraints on the cooling-flow model for those clusters.  The other clusters, however, show clear deviations from the cooling-flow model.

There are two significant residuals that should be noted.  The long wavelength region of the spectrum is not perfectly fit.  We believe this is due to errors in the effective area calibration, imperfect background modeling, and spatially broadened ($\sim$ 3 \AA\/) emission lines of O VIII, O VII, N VII, and C VI emission lines from the diffuse soft X-ray background (see \citealt{mccammon}).  These discrepancies have a minimal effect on our conclusions about the temperature distribution, however, since those rely primarily on the Fe L region.  Another problem is evident in the detailed line fitting of the spectrum of NGC 533.  This is probably due to an underprediction of the  2p-3s lines of Fe XVII and Fe XVIII (at 16 and 17 \AA\/, respectively).  There is a known problem with the excitation rates for these transitions \citep{beiersdorfer}, as has been discussed for the case of NGC 4636 by \cite{xu} and Capella by \cite{behar}.  

The differential luminosity distribution for all clusters is plotted in Figure 6.  The luminosity is normalized with respect to the prediction from the isobaric radiative multi-phase model using Equation (1) and $\dot{M}$ from Table 4.  The presence of points above the dotted line is not unexpected, since these can result from the background uncooled hot plasma.  This can also result from the ambiguity in defining a distinct cooling radius.  However, the data should closely track the line $y=1$ at lower temperatures, if the cooling-flow model is correct.  Many upper limits, however, are well below that line at temperatures near a third of the background temperatures.  As can be seen, the well-measured clusters show clear deviations from isothermality with significant plasma existing down to one-quarter of the background temperature.  In Figure 7, we plot the same results with the horizontal axis now scaled to the background temperature.  Here the discrepancies from the cooling-flow prediction look more systematic.  In particular, the differential luminosity distribution appears to be roughly consistent with the expression,

\begin{equation}
\frac{dL}{dT}= \frac{5}{2} \frac{ \dot{M} k}{\mu m_p} \left( \alpha+1 \right) {\left( \frac{T}{T_{0}} \right) }^{\alpha}
\end{equation}

\noindent
where $\alpha \approx 1$ to $2$ instead of 0, as expected for the isobaric radiative cooling-flow model. 
  There is still significant scatter in the normalization and slope of this relation, which could be due to various systematic effects, or to real differences between the cooling-flows.  Further detailed analyses that start with the assumption of Equation (3) might yield further insight into the scatter.  However, it is clear that no cluster has a temperature distribution consistent with the $\alpha=0$ case, and that in each case the failure of the cooling-flow model occurs at a fraction of the virial temperature rather than at a fixed value. 

The derived abundances as a function of ambient temperature are plotted in Figure 8.
The abundance of iron declines slightly with more massive clusters as indicated in earlier ASCA observations (\citealt{mushotzky2}, \citealt{fukazawa}), although the abundance measured here is somewhat higher.  The emission weighted iron abundance ranges from one third to 70$\%$ with higher values in lower mass clusters.  This presumably reflects the increase in stellar mass fraction for lower mass systems.  There are no obvious trends in the other abundance ratios and there is considerable scatter in the points.  This is possibly caused by subtle fitting biases.  In particular, the abundances reflect the emission-weighted abundance average in an aperature that varies from cluster to cluster and there is often an anti-correlation between the absolute abundances and the width of the differential emission measure distribution.  Additionally, abundances of elements other than iron are measured off the peak of their fractional ionic abundance, and depend sensitively on the temperature distribution.

Nevertheless, all of the abundances seem to cluster around particular ratios.  The oxygen to iron ratio is 70 $\pm$ 20 $\%$ (1 $\sigma$) of the solar value.  The magnesium to iron ratio is 100 $\pm$ 40 $\%$, the neon to iron ratio is 110 $\pm$ 40 $\%$, silicon to iron is 230 $\pm$ 80$\%$.  Abundance patterns for at least these four elements resemble those found in \cite{xu} suggesting a common enrichment history between ellipticals and clusters.  These particular ratios do not appear to fit into a simple superposition of Type Ia yields plus Type II supernovae integrated over a single initial mass function, as in \cite{gibson}.  They more closely resemble low-metallicity high-mass supernovae yields, but further analyses with more attention to additional elements are needed to establish the pattern completely and determine the spatial distribution.  Additionally, carbon and nitrogen abundances could be roughly estimated in M87 and Abell 496 to be $C/O$ $\sim$ 2 and $N/O$ $\sim$ 2.5 with large uncertainties.  This implies, however, a non-negligible fraction of chemical enrichment has been processed through stellar winds.  Further detailed analyses of the measured abundances and their spatial distributions will be presented elsewhere (\citealt{tamura3}).

The derived intervening column densities are generally consistent with the expected galactic values.  The only large deviation is with 2A0335+096, which has a high galactic column density and the measurements could be more affected by systematic uncertainties. In addition, allowing an intrinsic absorber with the standard cooling-flow model produced worse fits than the standard cooling-flow model alone.  Although the absorber can remove significant cold emission with column densities in excess of $2 \times 10^{21} \mbox{cm}^{-2}$, it also removes the low energy tail of the bremsstrahlung due to the core, which provides a significant fraction of the continuum.  Limits on intrisic absorbers are given in Table 4 and are generally less than $2 \times 10^{20} \mbox{cm}^{-2}$.  

\section{Discussion}

In \S7, we demonstrated the presence of significant quantities of plasma just below the ambient temperature, $T_0$, down to $T_0/2$ and the clear failure of the cooling-flow model at lower temperatures.  We also showed the consistency of the sample with the empirical differential luminosity distribution given by Equation (3) with the parameter, $\alpha$, equal to $\approx$ 1 to 2.  Here, we consider several ideas which have been suggested to explain these results.  We first discuss mechanisms which modify the isobaric cooling-flow model, but involve no real modification to the general paradigm of radiatively-driven flows in gravitationally-relaxed cluster cores.  Then, we discuss substantial modifications to the general cooling-flow process by inclusion of additional heating mechanisms or non X-ray cooling channels.

\subsection{Modifications to the Standard Spectral Prescription}

We have compared the data to the standard isobaric cooling-flow model.  A modification to this model is expected if a weak magnetic field is amplified as cooling plasma compresses in order to maintain pressure equilibrium.  At the lowest temperatures then, these blobs cool isochorically.  This would only modify the expression in Equation (1) by changing the $\case{5}{2}$ to $\case{3}{2}$, which is still inconsistent with the data displayed in Figures 6 and 7.

The compression of the gas due to the dark matter-dominated gravitational potential is included in our calculation of $\dot{M}$ by measuring directly both the temperature and density spatial gradients, and has generally been included in previous work (e.g. \citealt{allen2}, \citealt{peres}).  This is important if much of the gas is actually flowing, and the effect on the temperature distribution is given in \cite{nulsen2}.  It has a relatively small effect on the spectra apart from the normalization, however, and should not affect the luminosity at the lowest temperatures since most of the plasma is thought to drop out of the ambient medium before it flows (\citealt{johnstone}).  Even if there is no mass drop out, however, the $\case{5}{2}$ in Equation (1), would only be replaced by $\frac{3}{2}-{\lambda_T}/{\lambda_{\rho}}$ where $\lambda_T$ is the radial logarithmic temperature gradient and $\lambda_{\rho}$ is the radial logarithmic density gradient.  ${\lambda_T}/{\lambda_{\rho}}$ is always observed to be negative in the cores of clusters.  

Two recent explanations to the soft X-ray cooling-flow problem have been proposed that involve no real modification to the multi-phase model in Equation (1).  First, \cite{peterson1} suggested that differential absorption by colder material could absorb more of the emission from cold ions.  This is somewhat implausible, since one would expect hot and cold parcels of plasma to have roughly equal probability of being obscured.   Nevertheless, the result could be interpreted in the context of this model if the effective absorption column density scaled as a strong function of the occupied volume.  This would require the absorber to have an almost identical spatial distribution to the currently cooling parcels of gas from the cooling-flow and a small filling factor.  It is impossible to rule out this model in its most extreme form, since an infinite column density selectively applied to emission at a given temperature is, of course, identical to no emission at that temperature.  The most model-independent limits against this explanation comes from \cite{boehringer2} looking at absorption in spectra of AGN in clusters.

An explanation offered by \cite{fabian3} and developed in \cite{morris} involved a locally non-uniform distribution of metals in the ICM, which modifies the spectrum significantly.  This requires that the metal abundance variations occur on scales smaller than cooling parcels of plasma, and that the variations sample a particular bimodal distribution.  In its simplest form, this model would predict that the cooling-flow predictions would exceed the measured spectra near roughly a common temperature near 1 keV, where line cooling dominates over bremsstrahlung for super-solar abundances.  That is not what is observed, however, and this model is therefore not compatible with our results for the low temperature groups.

\subsection{Modifications to Radiatively-Driven Flows}

A variety of additional physical processes have been proposed to operate in the cores of clusters, which could, in principle, account for the soft X-ray cooling-flow problem (\citealt{peterson1}, \citealt{fabian3}, \citealt{boehringer2}, \citealt{sasaki}, and \citealt{brighenti} and references therein).   These include time-dependent heating by AGN outflows (e.g. \citealt{rosner}, \citealt{tabor}, \citealt{david}), electron thermal conduction from the outer regions of clusters (e.g. \citealt{tucker}, \citealt{stewart}, \citealt{bertschinger}, \citealt{voigt}, \citealt{zakamska}, \citealt{begelman2}), continual sub-cluster mergers (\citealt{markevitch2}), and interactions of the ICM with dark matter (\citealt{qin}).  Others invoke rapid cooling induced by mixing at interfaces with cold clouds (\citealt{begelman}, \citealt{fabian3}), dust mixing (\citealt{fabian3}), or acceleration of relativistic particles.

Most calculations have succeeded in establishing that the above mechanisms are energetically important, i.e these mechanisms have total heating powers that can roughly cancel radiative cooling energy losses, or that alternative cooling channels have sufficient luminosity to cool the plasma through other means.  The total energy balance (i.e. the sum of cooling luminosity minus heating power) determines the total mass deposited from the cooling-flow, and thus directly addresses the classic cooling-flow problem.  The total energy balance, however, only affects the soft X-ray cooling-flow problem by reducing the total normalization of Equation (3) by a fixed value.  It does not change the predicted differential luminosity distribution.

The critical difference, however, between the soft X-ray cooling-flow problem and the classic cooling-flow problem is that the latter requires a clear explanation for why X-ray cooling does not appear to be carried to completion.  It is difficult to find a relevant heating, mixing, or cooling time scale that would be comparable to the X-ray cooling time.   If the time scale for a given process is too short it will overwhelm the cooling-flow ($\alpha=\infty$), and if the time-scale is too long, it is dynamically unimportant ($\alpha=0$).
Below, we discuss the proposed physical processes and whether they can account, both energetically and dynamically, for the missing soft X-ray luminosity.

\subsubsection{External Heat Sources}

There are three requirements for additional heating mechanisms to be compatible with the observations.  First, the total time averaged heating power, $<$P$>$, has to roughly cancel the radiative losses so that the expression,

\begin{equation}
\frac{<P> \mu m_p}{ \frac{5}{2} k T_0 \dot{M}}
\end{equation}

\noindent
is near unity.  The denominator in the expression varies among the clusters in our sample by four orders of magnitude, so the proposed heating process must operate on a variety of scales.  Second, the heating has to be distributed spatially throughout the cooling-flow volume to cancel cooling everywhere.  Third, the process has to be self-regulating, so that the time scale for heating remains comparable to the cooling time for all clusters. 

{\bf AGN outflows:}  Time-dependent AGN outflow heating models have been considered by a variety of authors (e.g. \citealt{rosner}, \citealt{tabor}, \citealt{churazov}, \citealt{brueggen}, \citealt{quilis}, \citealt{david}, \citealt{nulsen3}).  Buoyant bubbles carrying relativistic plasma appear to be a common phenomena in clusters with central AGNs.  The thermal energy seems to be enough to heat cooling-flows through cosmic ray interactions and mechanical heating, but it is unclear whether this energy gets properly channeled into the cooling volume (e.g. \citealt{loewenstein}, \citealt{fabian3}).  Note that it is essential that the heat be distributed evenly throughout the region which is thermally unstable.  In addition these models must be made self-regulating to counteract cooling at a rate proportional to the mass deposition rate, and with periods of heating roughly as long as periods of cooling.  Clearly, it requires a significant degree of fine-tuning.

{\bf Electron Thermal Conduction to the Background Plasma:}  There is considerable thermal energy in the outer regions of clusters that can destroy any existing cooling-flow through electron thermal conduction (e.g. \citealt{tucker}, \citealt{stewart}, \citealt{bertschinger}).  The size of cooling-flows are only a few electron mean free paths in the absence of magnetic fields.  The critical question is to what level is conduction suppressed by tangled magnetic fields, an issue which continues to be debated theoretically (\citealt{chandran}, \citealt{narayan}).  Observationally, conduction is suppressed by factors near 100 in identified cold fronts (\citealt{ettori}, \citealt{markevitch}, \citealt{vikhlinin}).  \cite{voigt}, \cite{zakamska}, \cite{fabian5} have demonstrated that the heat flow from the outer regions of clusters with a small suppression ($>$ $10^{-1}$) in the Spitzer conductivity would appear to cancel radiative losses in many clusters.  The spatial distribution of the heating and overall energetic requirements appear to be satisfied by conduction models, but there is no explanation for why the cluster would cool to their current temperature distribution.  Since conduction suppresses temperature gradients by definition, this mechanism alone does not solve the dynamical problem presented here.

{\bf Sub-cluster Mergers:}  \cite{markevitch2} has shown that clusters previously thought to be fully relaxed exhibit temperature fronts consistent with plasma exhibiting large bulk motions in the gravitational potential.  This provides another source of energy that has not been dissipated and therefore leads to an increase of the cooling time above previous estimates (\citealt{gomez}), so that the time scales for radiative cooling and dynamical relaxation may be similar.  There is no reason for this process to be self-regulating, however, and we would expect a much larger difference in the temperature distributions, depending on each cluster's merger history.  This explanation also requires a conspiracy of factors to both cancel radiative cooling in global energetics, and ensure that the mergers occur with a frequency that allows some cooling.  Future numerical simulations may test this further, and presumably observations of cooling-flows at a different epoch would not show the same effects.

\subsubsection{Rapid Cooling Mechanisms (Heat Sinks)}

The radiative isobaric cooling-flow model assumes that all of the thermal energy is released in X-rays at high temperatures.  There may, however, be additional contributions from other cooling processes.  There are three main requirements for additional cooling channels to explain the observations, which are similar to, but slightly different from the heating requirements.  The first is that the total power in the coolant be comparable to the missing soft X-ray luminosity.  
Any coolant with power, $P_{coolant}$, reduces the total X-ray emission by a factor,

\begin{equation}
\frac{1}{1+\frac{P_{coolant}}{L_{x}}}
\end{equation}

\noindent
The second requirement is that the ratio $\frac{P_{coolant}}{L_{x}}$ needs to have a temperature dependence consistent with Equation (3) or that the cooling channel is self-regulating in the same sense as the discussed heating models.   The third requirement is that the energy should be released with a similar spatial distribution to the lowest temperature X-rays.  

{\bf UV cloud interfaces:}  \cite{begelman} and \cite{fabian3} discussed the possibility that hot electrons are cooled conductively by interfaces with cold clouds.  This leads to emission in the UV where the cooling function is the highest, and is consistent energetically with the large observed H$\alpha$ luminosities (\citealt{heckman}, \citealt{crawford}) of $10^{42}$ to $10^{44}$ ergs/s in cooling-flows.  However, by itself, this model does not explain the observed temperature distribution in the soft X-ray band, since high temperature electrons are cooled by this process as well.  In addition, in such a picture highly charged ions should also impact the cloud interfaces, resulting in charge exchange, which produces copious soft X-ray line emission.  The spatial distribution of H$\alpha$ is remarkably similar to the coolest X-rays (\citealt{ettori}) and the total luminosity is marginally sufficient to account for the missing soft X-ray luminosity (\citealt{fabian3}), but again the dynamical problem of partial cooling is not easily solved by this mechanism alone.

{\bf Dust Mixing:}  Some clusters have been established as strong IR sources, typically of order $10^{43}$ to $10^{45}$ ergs/s \citep{edge2}.  \cite{fabian4} demonstrated that in all systems studied, the missing thermal energy in soft X-rays is more than accounted for by the IR luminosity.  Hot electrons impinging on dust grains could collisionally heat the grains and this in turn cools the hot electrons.  The cooling function for this process has been discussed in \cite{silk} and \cite{draine} and is 100 times larger than the X-ray cooling function for a typical ISM composition of dust grains.  As such, this explanation appears promising, but there is no strong temperature dependence to either the X-ray or dust cooling function that would systematically reduce the overall level of soft X-ray emission.  Presumably, the success of this process depends on magnetic field topologies and a steady-state process of destruction and creation of dust grains.  Suppression of conduction by magnetic structures in some regions could play a critical role in regulating both UV cooling and dust mixing, and fast magnetic reconnection could in turn realign those structures.  The dynamical problem remains unsolved here too, and if dust cooling is important for cluster X-ray plasmas, it creates an even larger cooling-flow problem.

{\bf Relativistic Particles and Alfv\'{e}n Waves:} The increase in magnetic field density as cooling blobs compress may allow for considerable energy to be released via Alfv\'{e}n Waves or mildly relativistic particles.  In fact, the steeper synchrotron indices and short particle lifetimes in cluster radio mini-haloes have been invoked to argue for reacceleration of particles in cooling-flows (e.g. \citealt{petrosian}, \citealt{gitti}).  This process should, however, only  liberate a small fraction of the thermal energy and it provides no natural explanation for the observed temperature distribution.  

\subsection{General Considerations and Future Work}

Implicit in our discussion above is that only one mechanism explains the observations.  Given that many of the above ideas are energetically viable, that may be unlikely.  It may be that the soft X-ray cooling-flow problem and the classic cooling-flow problem are solved by different solutions.  It may also be possible that a combination of processes could have different effects.  For example, heating cooling-flows both from the inside with an AGN and from the outside through conduction, results in a more stable heating method than each mechanism does individually (\citealt{begelman2}).

We have demonstrated that the soft X-ray cooling-flow problem is not an isolated phenomenon and appears on a range of scales.  The empirical characterization offered here may provide clues to the ultimate solution.   Clearly, the model for the multiphase distribution is inadequate.  More detailed comparisons between observations at X-ray and other wavelengths may more carefully address which of the physical mechanisms are important energetically.  Much more theoretical work is needed in addressing why any of the proposed mechanisms would dynamically conspire to produce the observed temperature distribution.  These descriptions, however, are just beginning to be developed and it remains a theoretical challenge either to cool keV plasma with little resulting soft X-ray radiation, or to balance radiative cooling both energetically and dynamically.

\acknowledgements
This work is based on observations obtained with XMM-Newton, an ESA science mission with instruments and contributions directly funded by ESA Member States and the US (NASA). Work on the RGS at Columbia University and U. C. Berkeley is supported by NASA.  The laboratory for Space Research, Utrecht is supported by NWO, the Netherlands Organization for Scientific Research.  JRP acknowledges many helpful conversations with other members of the Columbia team.

\onecolumn

\begin{deluxetable}{lllllll}
\tablecaption{Important line blends in Cluster Soft X-ray Spectra
\label{tab:lines}}
\tablenum{1}
\tablewidth{0pt}
\tablehead{\colhead{Ion} &\colhead{Wavelengths} & \colhead{Temperature} & & \colhead{Ion} &\colhead{Wavelengths} & \colhead{Temperature}}
\startdata
          & \AA\/ & keV  \\ \tableline
Fe XXIV  & 10.6, 11.2 & 0.9 $\rightarrow$ 4.0 & &  O VIII & 19.0, 16.0 & $>$0.2 \\
Fe XXIII & 11.0, 11.4 & 0.8 $\rightarrow$ 2.0 & & O VII & 21.6, 22.0  & 0.1 $\rightarrow$ 0.2  \\
    & 12.2 & & & Si XIV & 6.2 & $>$1.0  \\
Fe XXII  & 11.8, 12.2 &  0.6 $\rightarrow$ 1.5 & & Si XIII & 6.6, 6.7 & 0.2 $\rightarrow$ 1.0  \\
Fe XXI	 & 12.2, 12.8 & 0.5 $\rightarrow$ 1.0  & & Al XIII & 7.2 & $>$1.2 \\
Fe XX    & 12.8, 13.5 & 0.4 $\rightarrow$ 1.0  & & Al XII  & 7.8, 7.9 & 0.3 $\rightarrow$ 1.2 \\
Fe XIX   & 13.5, 12.8 & 0.3 $\rightarrow$ 0.9  & & Mg XII & 8.4  & $>$0.7\\
Fe XVIII & 14.2, 16.0 & 0.3 $\rightarrow$ 0.8 & & Mg XI  & 9.2, 9.3 & 0.1 $\rightarrow$ 0.6 \\
Fe XVII & 15.0, 17.1 & 0.2 $\rightarrow$ 0.6 & & Ne X & 12.2  & $>$0.4 \\
	& 15.3, 16.8 & &  & Ne IX & 13.5, 13.7 & 0.1 $\rightarrow$ 0.3  \\
N VII & 24.8  & $>$0.1 \\
C VI & 33.7  & $>$0.1 \\
\enddata
\end{deluxetable}

\begin{deluxetable}{lrrrrr}
\tablecaption{Basic Properties of the Sample
\label{tab:basic}}
\tablenum{2}
\tablewidth{0pt}
\tablehead{
\colhead{Cluster} &
\colhead{XMM Rev.} &
\colhead{Eff. Exposure} & \colhead{Redshift} & \colhead{Lum. Distance} &
\colhead{Ang. Scale} }
\startdata
           & 	  & ks & 	& Mpc  & kpc/arcsec \\ \tableline
A 1835     & 0101 & 36 & 0.2528 & 1598 & 6.2\\
A 665	   & 0242 & 20 & 0.1818 & 1042 & 4.3\\
A 1795     & 0100 & 40 & 0.0622 & 296  & 1.4\\ 
Hydra A    & 0183 & 38 & 0.0538 & 253  & 1.2\\
Ser 159-03 & 0077 & 38 & 0.0580 & 274  & 1.3\\ 
2A0335+096 & 0215 & 26 & 0.0347 & 158  & 0.7\\ 
A 4059     & 0176 & 54 & 0.0460 & 213  & 1.0\\ 
A 496	   & 0211 & 29 & 0.0328 & 149  & 0.7\\
MKW 3s     & 0129 & 39 & 0.0442 & 204  & 1.0\\ 
A 2052     & 0128 & 33 & 0.0353 & 161  & 0.8\\ 
A 262      & 0203 & 36 & 0.0163 & 71   & 0.3\\
A 1837     & 0200 & 50 & 0.0372 & 170  & 0.8\\ 
M87        & 0097 & 42 & 0.0043 & 19   & 0.1\\
NGC 533    & 0195 & 48 & 0.0185 & 82   & 0.4\\  

\enddata
\end{deluxetable}

\begin{deluxetable}{lrrr}
\tablecaption{Spatial Parameters}
\label{tab:spatial}
\tablenum{3}
\tablewidth{0pt}
\tablehead{ \colhead{Cluster} & \colhead{$\beta$} & \colhead{${ r_{core}}$} 
& \colhead{${ r_{cool}}$} }
\startdata
          &  & arcsec &  arcsec \\ \tableline
A 1835    & 0.86 & 79  & 25  \\ 
A 665	  & 0.67 & 53  & 55      \\
A 1795    & 0.74 & 147  & 55 \\ 
Hydra A   & 0.73 & 138 & 44 \\
Ser 159-03& 0.79 & 72  & 29 \\ 
2A0335+096& 0.79 & 156 & 63    \\
A 4059    & 0.78 &  72 & 156 \\
A 496	  & 0.64 & 380 & 72  \\
MKW 3s    & 0.72 & 111 &  25   \\
A 2052    & 0.67 & 258 & 51    \\
A 262     & 0.51 & 163 & 31  \\
A 1837    & 0.67 & 134 & 12    \\
M87       & 0.51 & 178 & 35  \\
NGC 533   & 0.58 & 119 & 18   \\

\enddata
\end{deluxetable}

\begin{deluxetable}{lrrrrrrrr}
\tablecaption{Abundances and Absorption Column Densities
\label{tab:abund}}
\tablenum{4}
\tablewidth{0pt}
\tablehead{
\colhead{Cluster} &
\colhead{${ N_H}$} &
\colhead{${ N_H^{Galactic}}$}  & \colhead{$ N_H^{Intrinsic}$} & \colhead{O} & \colhead{Ne} & \colhead{Mg} & \colhead{Si} & \colhead{Fe}}
\startdata
           & $10^{20} {\rm cm}^{-2}$  &$10^{20} {\rm cm}^{-2}$   &$10^{20} {\rm cm}^{-2}$  & & \\

& &  &\\ \tableline
A 1835     & 3$\pm$0.5 & 2.32 & $<$6  & 0.19$\pm$0.08 & \ldots & \ldots & \ldots & 0.27  \\
A 665	   & 9$\pm$3.0 & 4.24 & \ldots & \ldots & \ldots & \ldots & \ldots & 0.33\\
A 1795     & 1$\pm$0.5 & 1.19 & $<$2  & 0.38$\pm$0.11 & 0.80$\pm$0.20 & 0.75$\pm$0.25 & 1.8$\pm$0.6 & 0.42$\pm$0.04\\ 
Hydra A    & 4$\pm$0.5 & 4.94 & $<$2 & 0.34$\pm$0.06 & 0.36$\pm$0.12 & 0.26$\pm$0.13 & 1.0$\pm$0.5 & 0.49$\pm$0.08 \\
Ser 159-03 & 0.5$\pm$0.5 & 1.79 & $<$2 & 0.20$\pm$0.07 & 0.54$\pm$0.09 & 0.41$\pm$0.19 & 1.4$\pm$0.3 & 0.55$\pm$0.05\\ 
2A0335+096 & 28$\pm$0.5 & 17.6 & $<$25 & 0.53$\pm$0.06 & 0.62$\pm$0.10 & 0.50$\pm$0.18 & 0.9$\pm$0.2 & 0.56$\pm$0.05 \\
A 4059     & 2$\pm$0.5 & 1.10 &  $<$4 & 0.52$\pm$0.14 & 0.70$\pm$0.18 & 0.98$\pm$0.22 & 1.9$\pm$0.5 & 0.66$\pm$0.07 \\ 
A 496	   & 8$\pm$0.5 & 4.58 & $<$5 & 0.44$\pm$0.07 & 0.70$\pm$0.22 & 0.68$\pm$0.25 & 1.9$\pm$0.5 & 0.78$\pm$0.08 \\
MKW 3s     & 2.5$\pm$0.5 & 3.03 & $<$2 & 0.35$\pm$0.06 & 0.60$\pm$0.16 & 0.63$\pm$0.16 & 0.6$\pm$0.4 & 0.58$\pm$0.06 \\ 
A 2052     & 3$\pm$0.5 & 2.73 & $<$3 & 0.46$\pm$0.10 & 0.51$\pm$0.15 & 0.84$\pm$0.28 & 1.4$\pm$0.4 & 0.61$\pm$0.06 \\ 
A 262      & 9.5$\pm$0.5 & 5.37 & $<$12 & 0.51$\pm$0.09 & 0.37$\pm$0.22 & 0.87$\pm$0.23 & 1.8$\pm$0.4 & 0.68$\pm$0.07\\
A 1837     & 4$\pm$0.5 & 4.30 & \ldots & \ldots & \ldots & \ldots & \ldots & \ldots \\
M87        & 4$\pm$0.5 & 2.54 & $<$2 &0.40$\pm$0.04 & 0.70$\pm$0.07 & 0.55$\pm$0.06 & 0.9$\pm$0.1 & 0.54$\pm$0.05 \\
NGC 533    & 3.10 & 3.10 &  $<$18 & 0.41$\pm$0.05 & 0.52$\pm$0.25 & 0.98$\pm$0.30 & 2.0$\pm$0.5 & 0.89$\pm$0.15 \\  
\enddata
\end{deluxetable}

\begin{deluxetable}{lrrrrrrrrr}
\tablecaption{Cooling Luminosity}
\label{tab:dem}
\tablenum{5}
\tablewidth{0pt}
\tablehead{
\colhead{Cluster} &
 ${\dot{M}_{{\frac{1}{16} \leftarrow \frac{1}{8}}}}$ & ${\dot{M}_{\frac{1}{8} \leftarrow \frac{1}{4}}}$ & ${\dot{M}_{\frac{1}{4} \leftarrow \frac{1}{2}}}$ &
${\dot{M}_{\frac{1}{2} \leftarrow{1}}}$ & ${ L_x^{isothermal}}$ & $kT_{0}$ &
${\dot{M}_{morphological}}$ }
\startdata
 & ${\rm M_{\sun}} {\rm yr}^{-1}$& ${\rm M_{\sun}} {\rm yr}^{-1}$& ${\rm M_{\sun}} {\rm yr}^{-1}$ & ${\rm M_{\sun}} {\rm yr}^{-1}$& ${\rm ergs~s^{-1}}$& {\rm keV} &${\rm M_{\sun}} {\rm yr}^{-1}$  \\ \hline
A 1835     & $<$200 & $<$300 & 800$\pm$200 & 5800$\pm$800  & 6$\times 10^{45}$ & 9.5$\pm$0.5 & 1000  \\
A 665	   &  \ldots & \ldots & \ldots & \ldots & 2$\times10^{45}$ & 7.7$\pm$0.5 & \ldots \\
A 1795     & $<$30 & $<$30 & $<$80 & 380$\pm$40   & 2$\times 10^{45}$ &  5.5$\pm$0.5 & 300 \\ 
Hydra A    & $<$100 & 35$\pm$20 & 120$\pm$60 & 180$\pm$50 & 7$\times10^{44}$ & 6.0$\pm$0.3 & 180 \\
 Ser 159-03  & $<$30 & $<$30 & $<$60 & 210$\pm$30 & 3$\times10^{44}$ & 3.8$\pm$0.3 & 79 \\
2A0335+096 & $<$84 & 20$\pm$10 & 40$\pm$20 & 420$\pm$50 & 4$\times10^{44}$ & 3.2$\pm$0.3 & 170 \\
A 4059     & $<$10 & 10$\pm$5 & 40$\pm$20 & 100$\pm$10 & 3$\times10^{44}$& 6.0$\pm$0.3 & 60 \\ 
A 496	   & $<$10 & $<$15 & 25$\pm$10 & 120$\pm$20 & 8$\times10^{44}$& 4.7$\pm$0.3 & 72 \\
MKW 3s     & $<$64 & $<$10 & $<$20 & 45$\pm$10 & 3$\times10^{44}$& 3.7$\pm$0.3 & 13 \\ 
A 2052     & $<$10 & $<$10 & 15$\pm$5 & 100$\pm$20  & 4$\times10^{44}$& 3.4$\pm$0.3 & 50\\ 
A 262      & $<$10 & $<$2 & 1.6$\pm$0.5 & 10$\pm$1 & 9$\times10^{43}$& 2.1$\pm$0.2 & 2.0\\
A 1837     & \ldots & \ldots & \ldots &  \ldots  & 8$\times10^{43}$ & 4.2$\pm$0.4 & \ldots \\ 
M87  &  $<$12 & $<$0.6 & 0.6$\pm$0.2 & 5.9$\pm$0.5 & 4$\times10^{43}$ & 2.0$\pm$0.1 & 2.4 \\
NGC 533    & $<$5 & $<$1.6 & 2.3$\pm$0.4 & 5$\pm$0.5 & 6$\times10^{42}$ & 1.5$\pm$0.1 & 1.5 & \\  
\enddata
\end{deluxetable}

\clearpage

\begin{figure}
\includegraphics[width=7.00in]{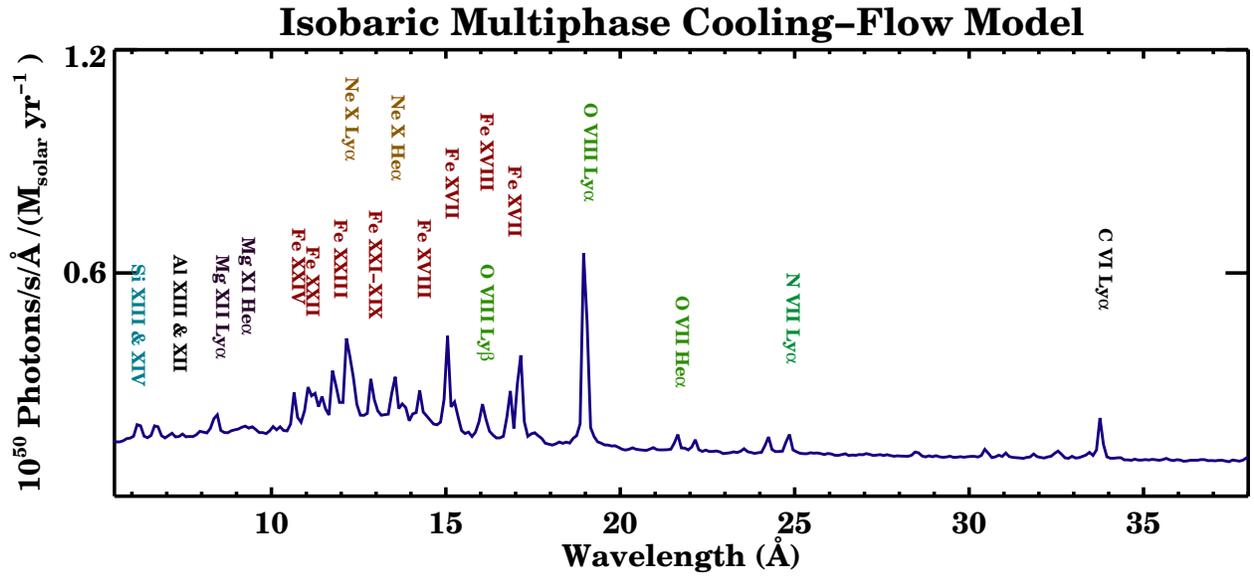}
\caption{The predicted spectrum of the isobaric multiphase model for a maximum temperature of 6 keV.  Note that Fe L lines and O VIII are particularly prominent.}
\end{figure}

\begin{figure}
\includegraphics[width=7.00in]{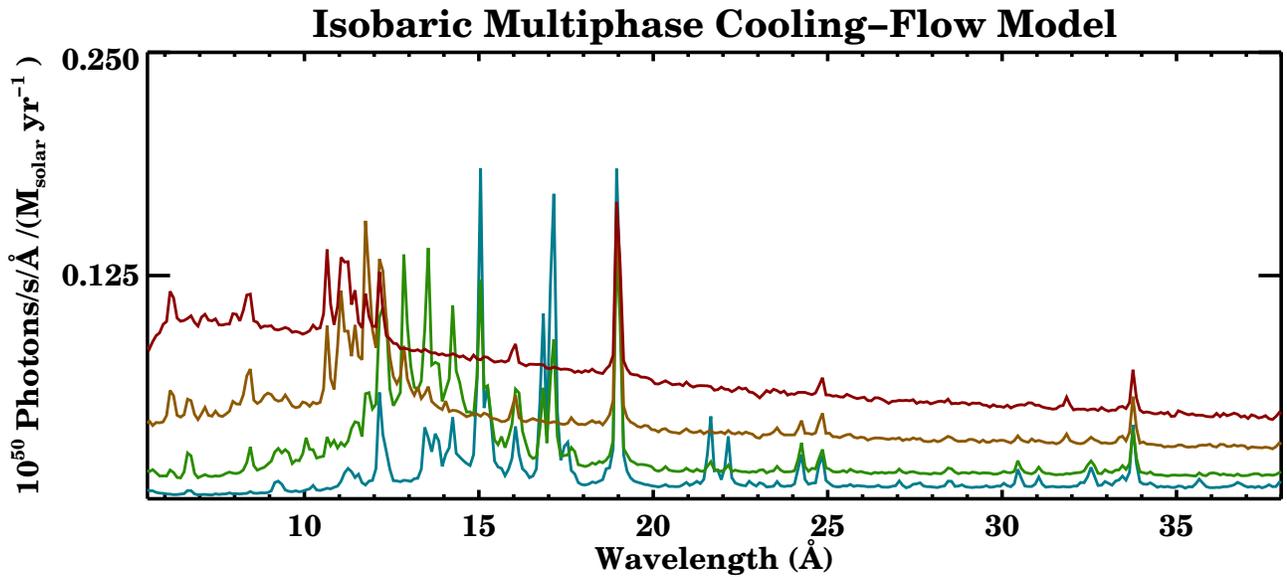}
\caption{The same plot as above but divided into different temperature ranges:  6 to 3 keV (red), 3 to 1.5 keV (yellow), 1.5 to 0.75 keV (green), and 0.75 to 0.375 keV (blue).  Note that the Fe L ions provide critical diagnostics of the temperature distribution; whereas, the Ly$\alpha$ transition from hydrogenic ions is produced at a large range in temperatures.}
\end{figure}

\clearpage

\includegraphics[width=7.00in]{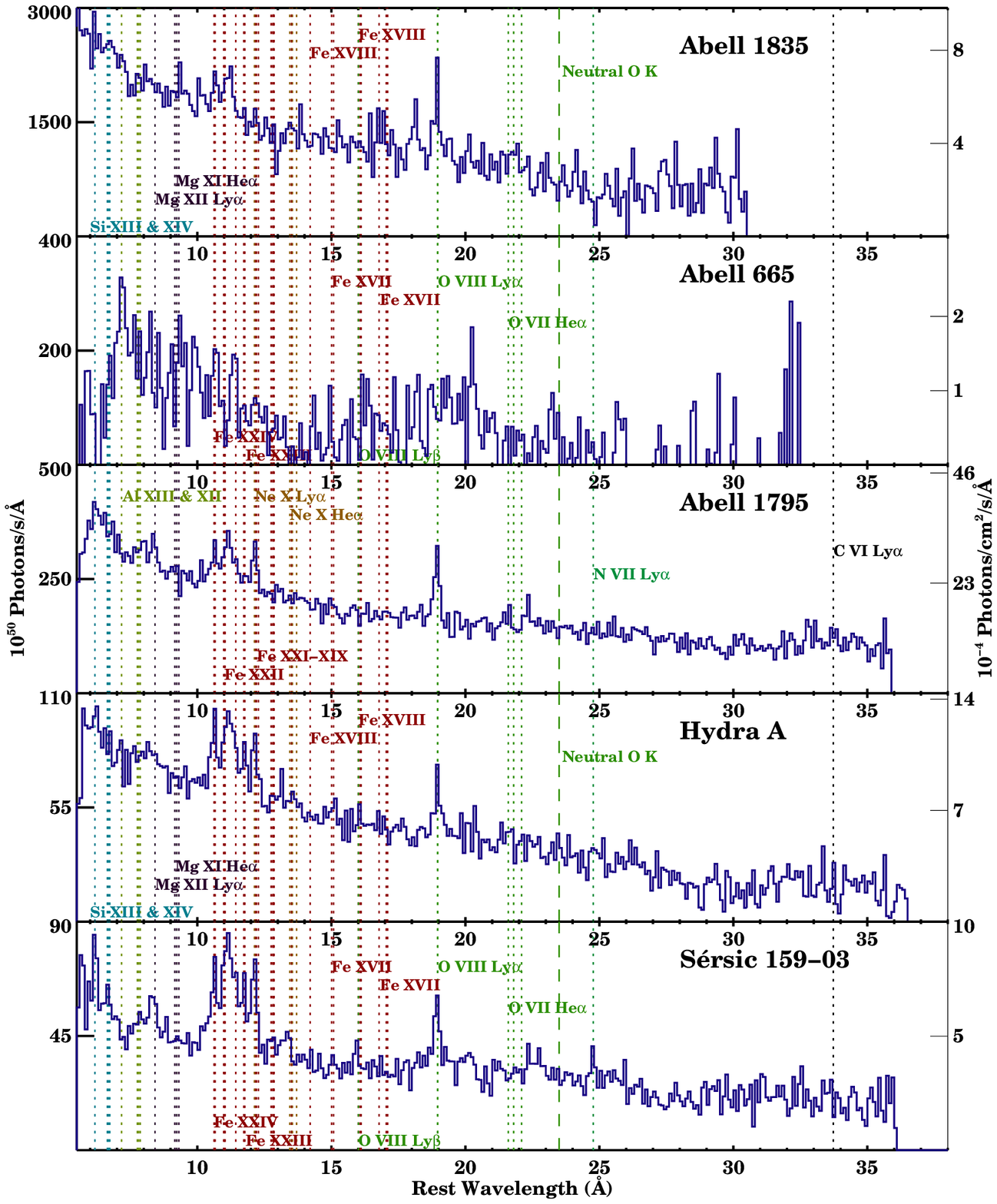}
\clearpage
\includegraphics[width=7.00in]{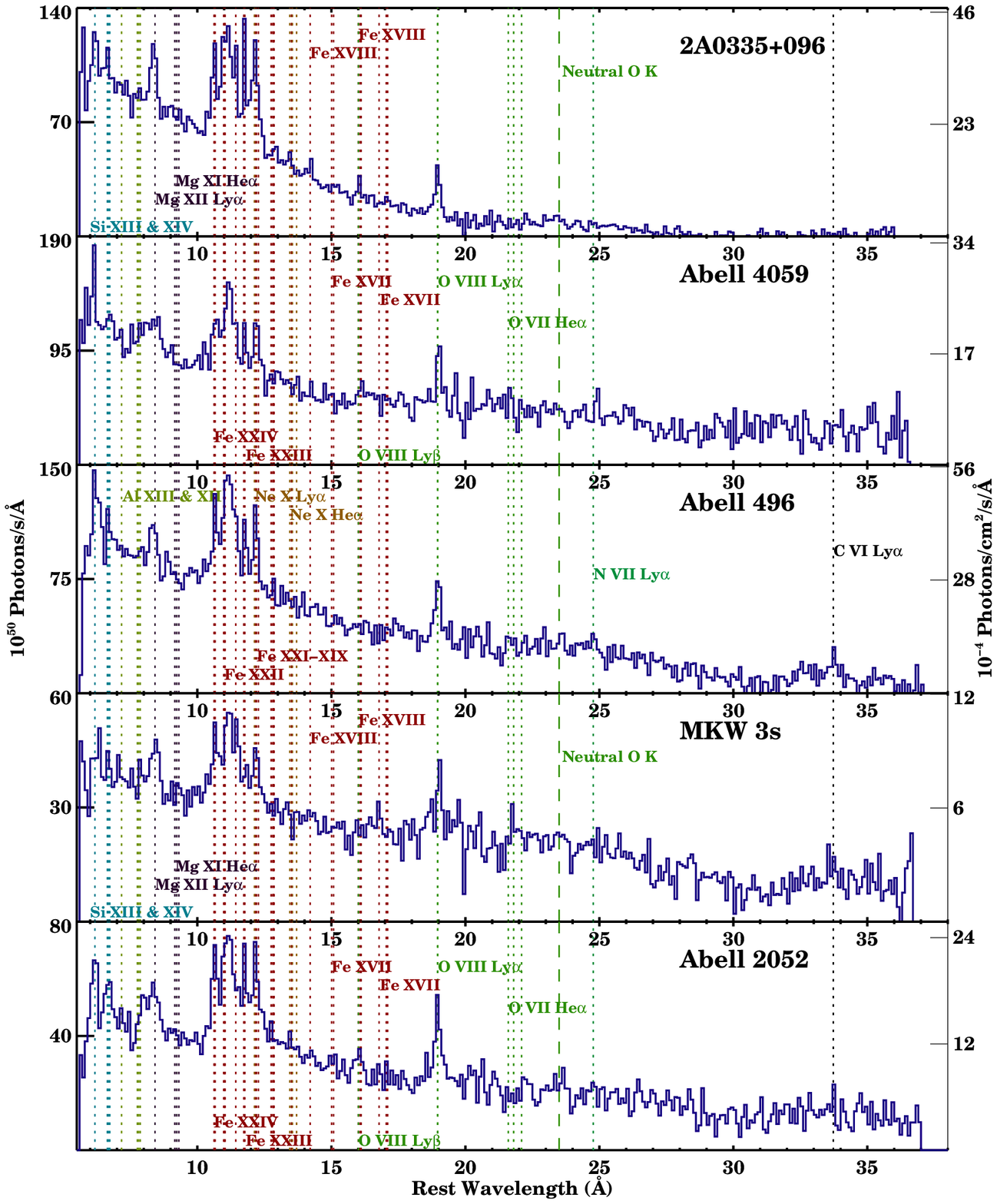}
\clearpage
\begin{figure}
\includegraphics[width=7.00in]{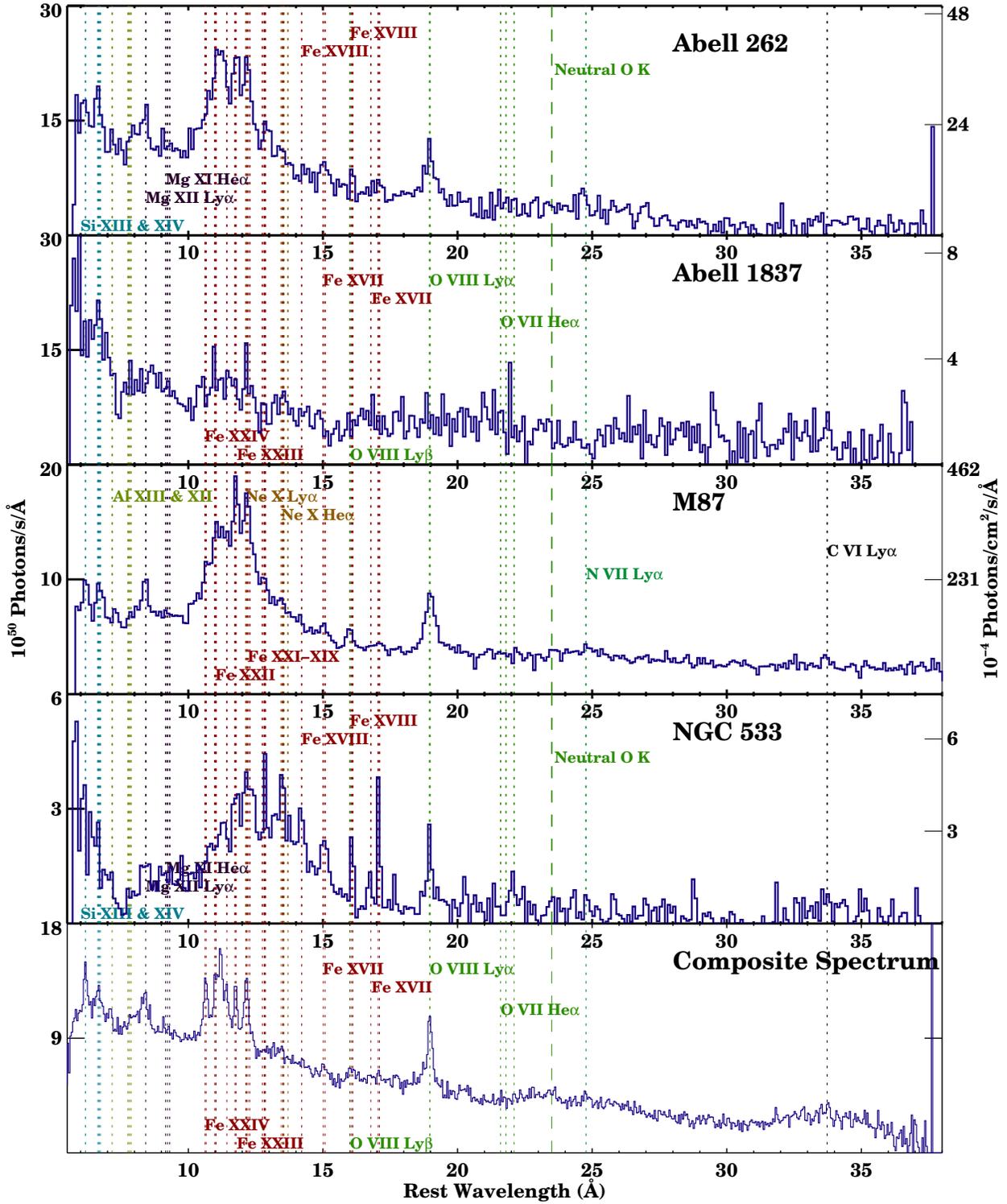}
\caption{(Three panels) Fluxed and background subtracted spectra of the sample of clusters.  Both RGS instruments and both spectral orders are included in all the plots.  Prominent detected or expected emission lines are labeled (also see Table 1) as well as the neutral O K edge.  All spectra are deredshifted so the horizontal axis is the wavelength in the cluster rest frame.  The last panel shows all the cluster spectra added together with the exception of M87.  The spectra are divided by the effective area so some regions of the spectrum contain more statistical noise than others.  In particular, the long wavelength region and the regions between 20 and 24 \AA\/ and 10.5 and 14 \AA\/ (in the lab frame) have lower effective area.}
\end{figure}

\clearpage
\includegraphics[width=7.00in]{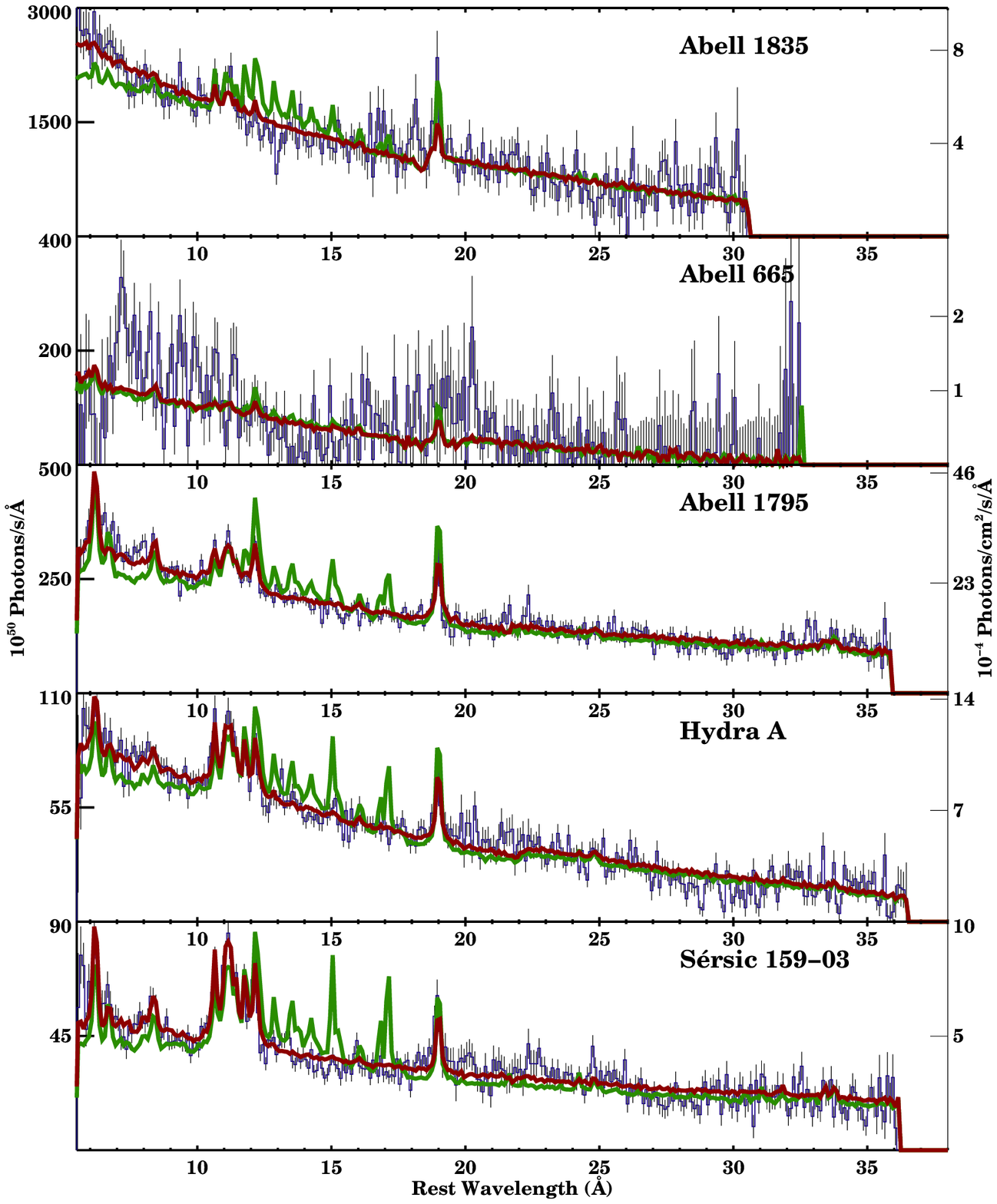}
\clearpage
\includegraphics[width=7.00in]{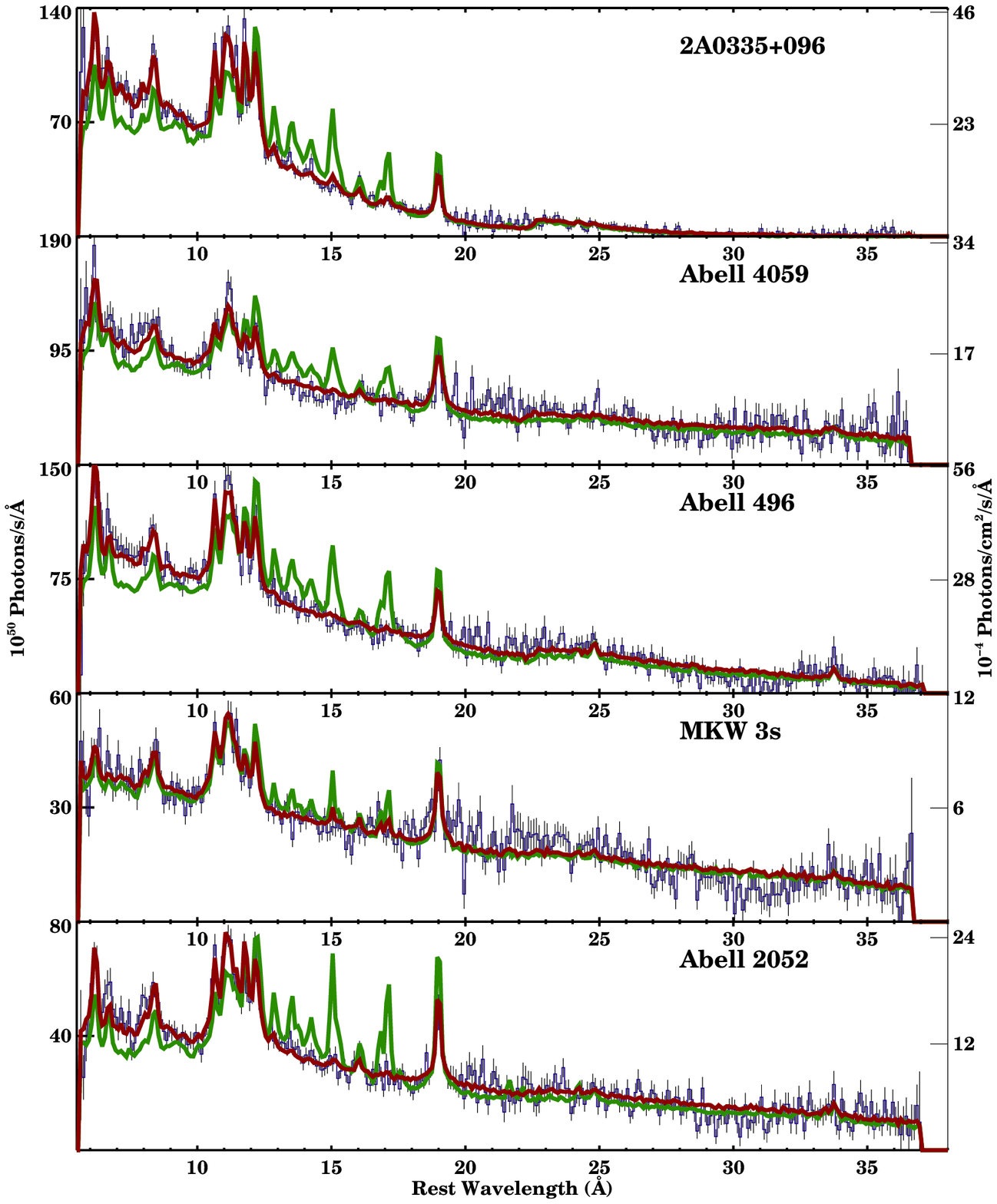}
\clearpage
\begin{figure}
\includegraphics[width=7.00in]{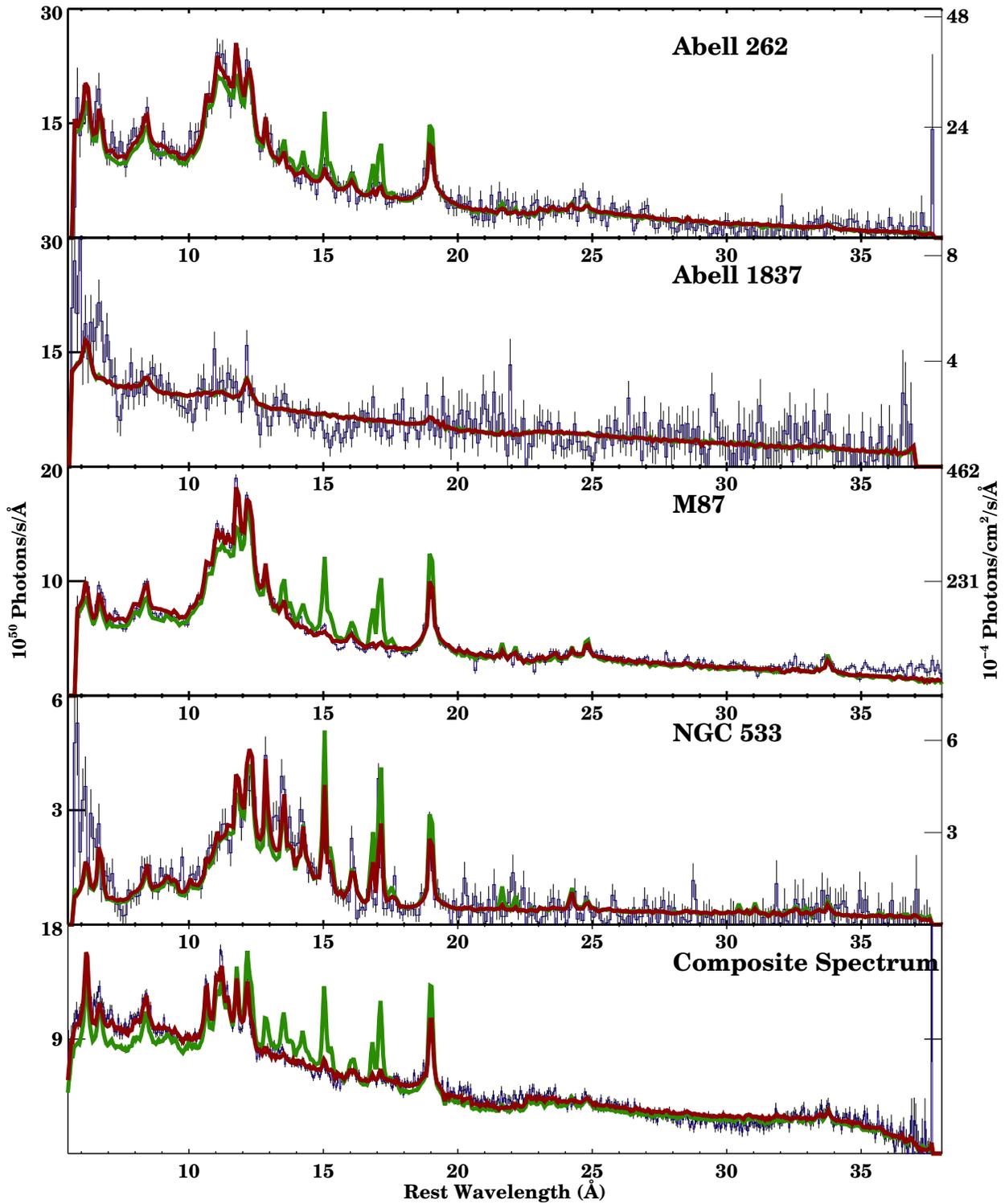}
\caption{(Three panels) Comparison of the data (blue) with 1 $\sigma$ error bars, the empirical best fit model (red), and the standard cooling-flow model (green).  Both the model and the data are fluxed and background subtracted.  The cooling-flow model is not a best fit, but calculated by merely taking the soft X-ray flux in the empirical model and using the standard isobaric temperature distribution.   Evident in all spectra is the severe overprediction of emission lines from the lowest temperatures.}
\end{figure}

\clearpage

\begin{figure}
\includegraphics[width=6.75in]{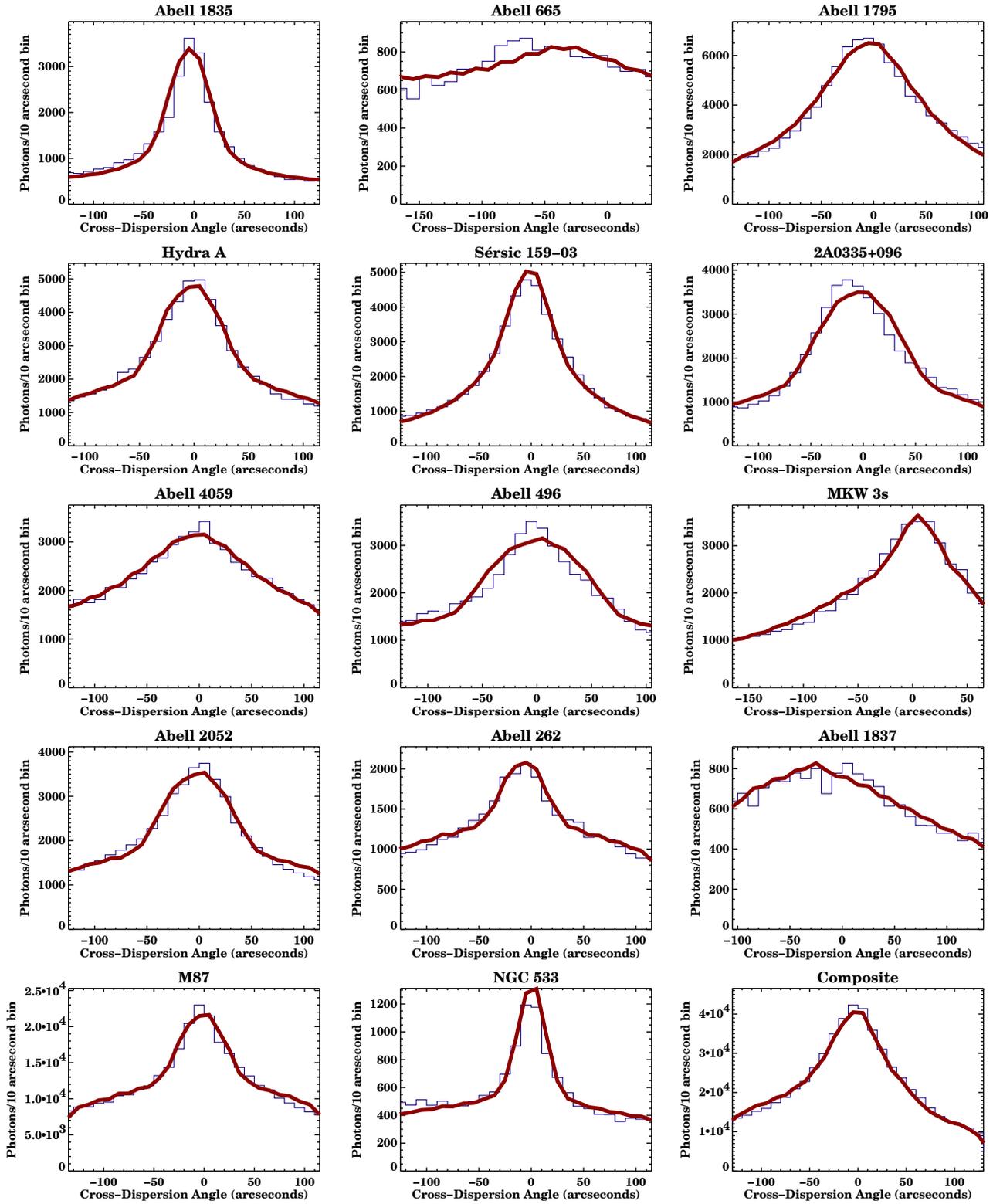}
\caption{Cross-dispersion profiles (1d images) of each individual cluster and all profiles added together in the composite spectrum.  The data are shown as a blue histogram and the model is the red line.  Most discrepancies seem to result from the assumption of spherical symmetry in the model.}
\end{figure}

\clearpage

\begin{figure}
\includegraphics[width=6.75in]{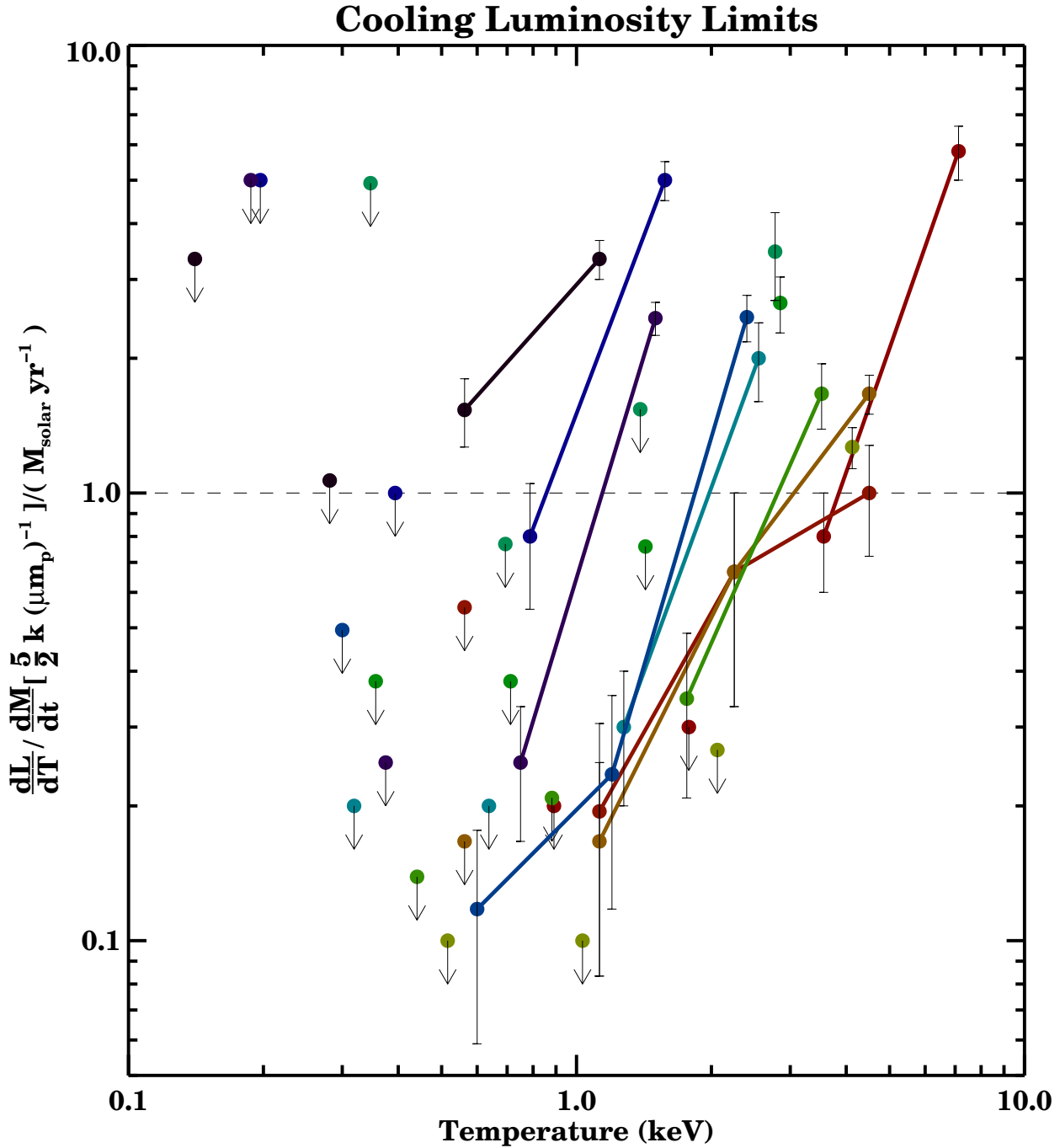}
\caption{Differential Luminosity vs. Temperature computed by dividing the upper limit on $\dot{M}$ by the estimate derived from the imaging data.  Each color represents a different cluster and lines connect the points, which are not upper limits. Points above the line are consistent with ambient plasma which has not cooled, but the isobaric multiphase model would predict that emission should closely track the dashed line at $y=1$ or the line $y=\frac{3}{5}$ for isochoric cooling.  A large number of upper limits and points lie below those lines, however.  The spectra generally become inconsistent with the model near $\frac{1}{3}$ of the ambient temperature.}
\end{figure}

\clearpage

\begin{figure}
\includegraphics[width=6.75in]{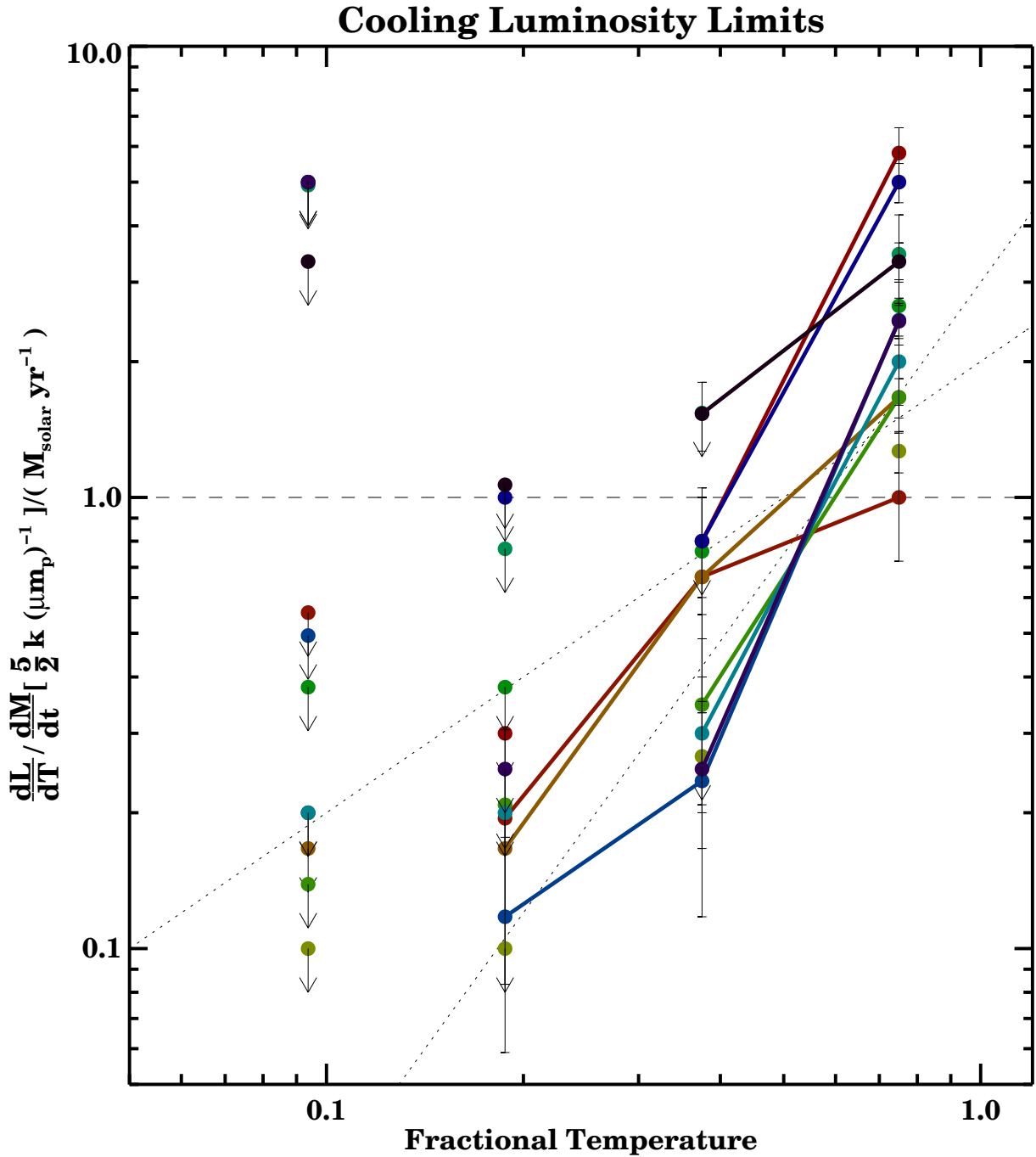}
\caption{The same as the previous figure except plotted as function of the fraction of the ambient temperature.  A general trend in the points which are not upper limits can be seen which follows a temperature distribution which is proportional to the fractional temperature or the fractional temperature to the second power (dotted lines).}
\end{figure}

\clearpage

\begin{figure}
\includegraphics[width=6.75in]{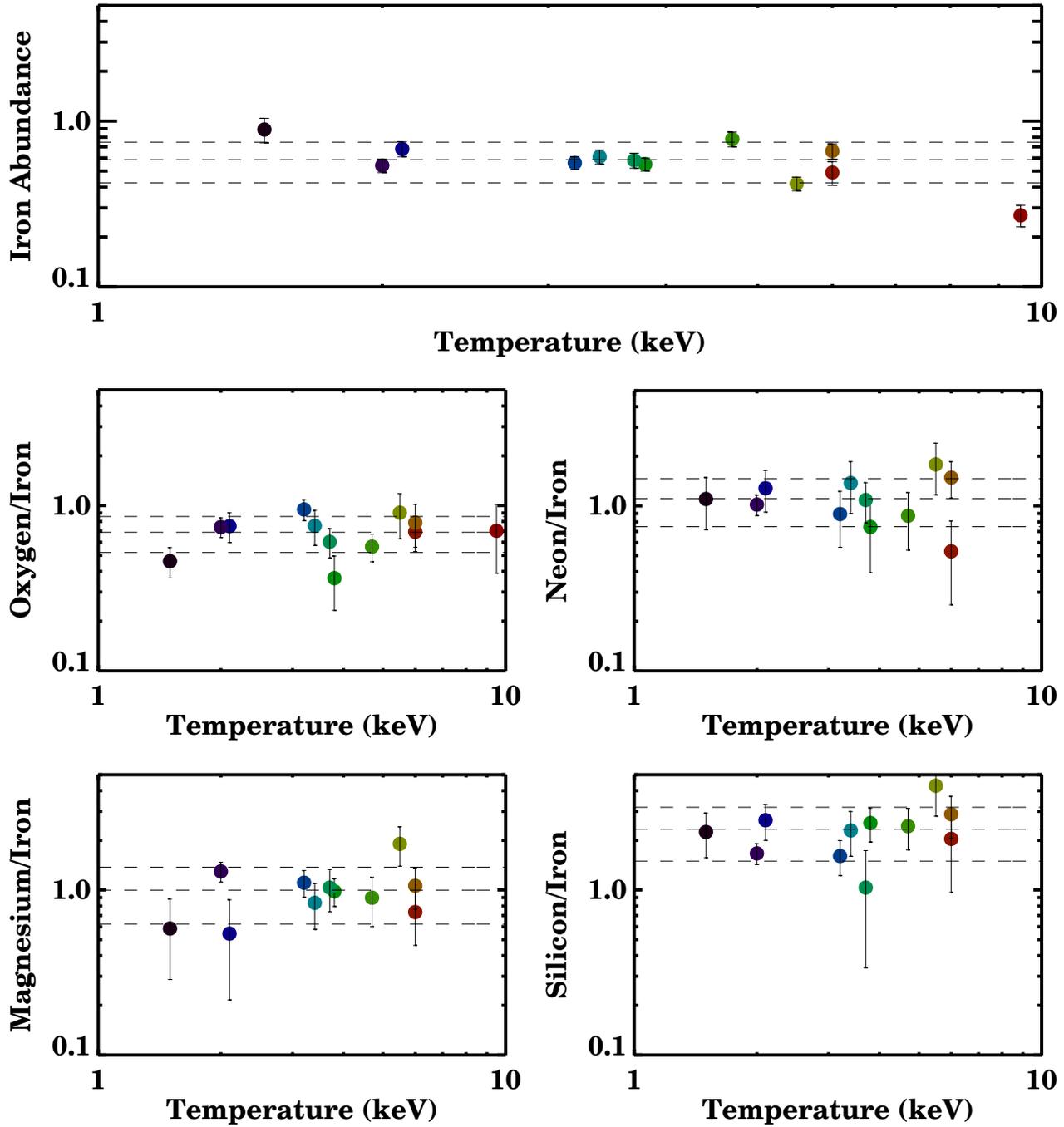}
\caption{Abundances and abundance ratios plotted at a function of the ambient cluster temperature.  The iron abundance tends to increase for lower mass systems.  No other obvious trends can be seen.  The dashed lines are the average value for the sample and the 1 $\sigma$ ranges.}
\end{figure}

\end{document}